\documentclass{article}
\usepackage[utf8]{inputenc}
\usepackage{amsmath}
\usepackage{amssymb}
\usepackage{graphicx}
\usepackage{enumerate}
\usepackage[square,sort,comma,numbers]{natbib}
\usepackage{xcolor}
\usepackage{authblk}
\usepackage[normalem]{ulem} 

\begin{document}
\title{Modestly weighted logrank tests}
\author[1]{Dominic Magirr} 
\author[1]{Carl-Fredrik Burman}
\affil[1]{Advanced Analytics Centre, AstraZeneca.}

\maketitle

\begin{abstract}

We propose a new class of weighted logrank tests (WLRT) that control the risk of concluding that a new drug is more efficacious than standard of care, when, in fact, it is uniformly inferior. Perhaps surprisingly, this risk is not controlled for WLRT in general. Tests from this new class can be constructed to have high power under a delayed-onset treatment effect scenario, as well as being almost as efficient as the standard logrank test under proportional hazards.

\end{abstract}


\section{Introduction}
The common way to analyze two-arm randomised controlled trials (RCTs) with a time-to-event endpoint is using a logrank test (LRT) or Cox proportional hazards model. The LRT is the most powerful rank-invariant test under the proportional hazards assumption \citep{peto1972asymptotically}. However, data from several recent immuno-oncology trials have clearly violated this assumption, implying that statistical power may be reduced, and hazard ratios may be difficult to interpret. The typical pattern is that the two treatment arms have similar survival or progression-free survival (PFS) for an initial period (a number of months), followed by a divergence \citep{alexander2018hazards}. This suggests that there is scope to improve over LRT in these trials, and, given the importance of this class of drugs \citep{couzin2013cancer}, the search for alternative methods has become very active.

Alternatives tend to fall into one of two categories: procedures based on weighted log-rank tests (\cite{gares2014comparison,karrison2016versatile,lin2017estimation,yang2010improved}) building on the work of \cite{fleming1981class}, or procedures based on weighted differences in survival curves (\cite{royston2013restricted,royston2016augmenting,tian2017efficiency,uno2014moving,shen2001maximum}) building on the work of \cite{pepe1989weighted}. The latter category includes restricted-mean survival time (RMST) and landmark analysis as special cases.

This paper is about weighted log-rank tests (WLRT). First, we show that previously proposed WLRTs may, if misinterpreted, lead to an increased risk of erroneously concluding that a new drug has better efficacy than standard of care. Second, based on results from \cite{leton2001equivalence}, we find specific choices of weights such that this risk is always controlled. In a simulation study, we compare the power of this new class of WLRT with the standard logrank test, previously proposed WLRTs, and a landmark analysis.

\section{Weighted logrank tests}

Let $S_0(t)$ and $S_1(t)$ denote the survival probability at time $t\geq 0$ on the control and experimental treatment arms, respectively. We make a distinction between a `weak' null hypothesis that the survival distributions are the same:

$$H_0^{\text{weak}}: S_1(t) = S_0(t) \text{ for all } t$$ 

and a `strong' null hypothesis that survival on the experimental arm is stochastically less than or equal to survival on the control arm:

$$H_0^{\text{strong}}: S_1(t) \leq S_0(t) \text{ for all } t.$$

Let $t_1<\ldots<t_k$ denote the $k$ distinct, ordered event times. The number at risk at time $t_j$ on arm $i=0,1$ is denoted by $n_{i,j}$, with $n_j := n_{0,j} + n_{1,j}$. The number of events on arm $i=0,1$  at time $t_j$ is denoted by $d_{i,j}$, with $d_j := d_{0,j} + d_{1,j}$. For weights $w_1,\ldots,w_k$, the weighted logrank test statistic is

\begin{equation}\label{logrankU}
U =  \sum_{j = 1}^k w_j \left( d_{0,j} - d_j \frac{n_{0,j}}{n_j} \right).
\end{equation}

Assuming that censoring is independent in each treatment group, under $H_0^{\text{weak}}$, $U$ is asymptotically normally distributed with 

$$E(U) = 0 \text{ and }\hat{V}(U) = \sum_{j = 1}^k w_j^2 \frac{n_{0,j}n_{1,j}d_j(n_j - d_j)}{n_j^2(n_j - 1)},$$

see, e.g., Chapter 7 of \citep{klein2005survival}. Statistical papers usually state that if $U^2 / V$ is larger than the $(1-\alpha)$-quantile of a one degree of freedom chi-square distribution, one may reject $H_0^{\text{weak}}$, and that this procedure has type I error $\alpha$. However, rejection of the weak null indicates nothing about the nature of the difference in survival functions between the two groups.

The logrank test is routinely accompanied by a point estimate and a confidence interval for the hazard ratio; for some estimators see \cite{berry1991comparison}. When the estimated HR and the confidence interval is below $1$, it is natural to conclude that the experimental treatment is superior.

\begin{figure}[htbp]
\centering
\includegraphics[width=\linewidth]{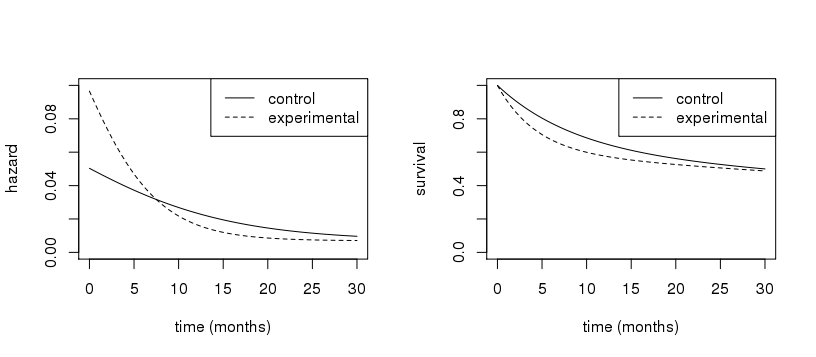}
\caption{An example where survival on the experimental arm is stochastically uniformly worse than on the control arm, but where the hazard functions cross, such that some WLRT will not control the type I error rate under $H_0^{\text{strong}}$.}
\label{hazard_survival}
\end{figure}

The WLRT that have been proposed recently in immuno-oncology, for example  \cite{he2015novel, lin2017estimation, xu2017designing}, have small $w_j$ for early $t_j$, where we expect little difference between survival curves, and larger $w_j$ for later $t_j$. "Power" is defined as $\text{pr}(U/\sqrt(V) > \Phi^{-1}(1 - \alpha / 2))$, and the new WLRT are shown, in delayed effect scenarios, to be far superior to standard LRT using this metric. Here, we must take great caution. Figure \ref{hazard_survival} shows a scenario of a 2-arm RCT where the survival distributions satisfy $H_0^{\text{strong}}$, with the experimental treatment being uniformly worse than control, but the hazard functions for the two groups are crossing. If the weights used were $w_j = 0$ for $t_j < 6$ (where the hazard ratio favours control) and $w_j = 1$ for $t_j \geq 6$ (where the hazard ratio favours experimental) then it is clear that $E(U)$ would be strictly positive and increasing with the sample size (number of events). If we were to claim that the experimental treatment is more efficacious than the control when $U/\sqrt(V) > \Phi^{-1}(1 - \alpha / 2)$, our risk of false positives may far exceed $\alpha / 2$.

\section{Strong type I error control}

\subsection{Is the standard logrank test really unweighted?}
Obviously, we want survival to be as long as possible. To test the strong null hypothesis, any reasonable test should favour longer survival over shorter survival. Say that two random patients in the control and experimental arms, respectively, have survival times $T$ and $T^\prime$. If the patient on experimental treatment lives longer, $T^\prime>T$, it is an indication of a treatment benefit. If the survival times of the two patients would be switched, we would certainly have \textit{less} reason to reject the strong null. Consequently, the p-value should increase. Many proposed weighted logrank tests violate this principle. 

To build intuition, we start by dissecting the standard logrank test, in a simplified situation. Consider a two-arm trial with $n$ patients per arm where $k$ events are observed prior to any censoring, but with the remaining $2n - k$ observations censored. Also assume that there are no tied observations and let $l_0$ denote the number of censored observations on the control arm. \cite{leton2001equivalence} show that the standard logrank statistic $U$ can be written as a weighted sum of the treatment arm indicators, $d_{0,j}$, minus a constant term, $K>0$, times the number of censored observations on the control arm:

$$ \sum_{j=1}^kd_{0,j} \left( 1 - \sum_{i=1}^j \frac{1}{2n - (i-1)} \right) - K \times l_0.$$

\begin{figure}[htbp]
\centering
\includegraphics[width=\linewidth]{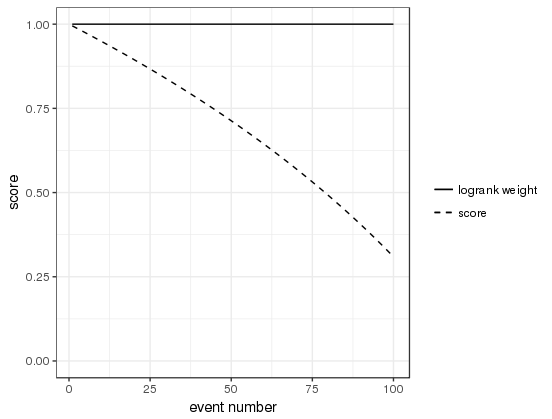}
\caption{Score-test representation of the standard logrank test where, with 100 patients per arm, there are 100 events prior to 100 censored observations.}
\label{lr_scores}
\end{figure}

The weights are more commonly known as "scores" in this context, and are decreasing in $j$. Figure \ref{lr_scores} shows an example with $2\, n=200$ and $k = 100$. 

 If an alternative test statistic of this form were to be proposed,

$$ \sum_{j=1}^kd_{0,j} c_j - K^{\prime} \times l_0,$$

but where the scores $c_j$ sometimes increased, so that $c_{j_1}<c_{j_2}$ when $j_1<j_2$, then the test would consider it "better" to have an earlier event number $j_1$ than a later $j_2$. We think such tests should normally not be accepted\footnote{It is possible to construct a set of scores that are not uniformly non-increasing and still lead to strong alpha control, but it would not make sense to use this.}. 
However, it could be possible to keep the scores constant. In fact, this would lead to a landmark analysis, where we compare the total number of events on each arm up to a certain time point. Our idea, which we will elaborate in subsequent sections, is to use constant scores $c_j = 1$ for the initial events, then, at a certain point in time, let $c_j$ decrease at a rate that is similar to the standard logrank test. Since we know that the standard logrank test is the optimal rank-preserving test under proportional hazards, our hope is that by imposing a limit on early scores (to avoid up-weighting events where no efficacy is expected), we can find a suitable test for a delayed-onset treatment effect.

\subsection{Score statistics}

Leton and Zuluaga \cite{leton2001equivalence} describe the relationship between logrank and score statistics under more general censoring patterns. Let $l_{i,j}$ denote the number of censored observations on arm $i = 0,1$ between times $t_j$ and $t_{j + 1}$, with $l_j: = l_{0,j}+l_{1,j}$. For scores $c_1,\ldots,c_k$ and $C_1,\ldots,C_k$ with $\sum_{j = 1}^k d_j c_j + \sum_{j = 1}^k l_jC_j = 0$, we define the score statistic:

$$S = \sum_{j = 1}^k d_{0,j} c_j + \sum_{j = 1}^k l_{0,j} C_j.$$

Under $H_0^{\text{weak}}$, and assuming equal censoring distributions on the two arms, permutation arguments (see , e.g., \cite{hothorn2008implementing}) can be used to show that $S$ is asymptotically normally distributed with

$$E(S) = 0,$$

$$V(S) = \frac{N_0 N_1}{N(N - 1)} \left(  \sum_{j = 1}^k d_{j} c_j^2 + \sum_{j = 1}^k l_{j} C_j^2  \right).$$

Leton and Zuluaga \cite{leton2001equivalence} show that for certain $w_j$, $c_j$ and $C_j$ the score and logrank statistics, $S$ and $U$, are equivalent. The conditions on the $w_j$, $c_j$ and $C_j$ are:
\begin{equation}\label{w_to_c}
C_j = -\sum_{i = 1}^j w_i \frac{d_i}{n_i}\text{,}\quad w_j = c_j - C_j
\end{equation}
or, equivalently,
\begin{equation}\label{c_to_w}
w_{j + 1} = (w_j + c_{j+1} - c_j) \frac{n_{j+1}}{n_{j+1} - d_{j+1}}\text{,}\quad w_j = c_j - C_j.
\end{equation}

For the standard logrank statistic, where $w_j = 1$ for all $j$, we see from (\ref{w_to_c}) that the $c_j$ and $C_j$ are uniformly non-increasing. This observation provides a heuristic argument for the strong type I error control of the standard logrank test. Fix any survival function $S_0(t)$. First assume the weak null, that $S_1(t)=S_0(t)$ for all $t\geq0$, and let a realization of data have survival and censoring times $\{T^\prime_i\}$ and $\{\tau_i\}$ for the experimental arm. The permutation argument \citep{hothorn2008implementing} shows control of the Type I Error. Now consider the strong null, $S_1(t)\leq S_0(t)$ for all $t\geq0$. Using a coupling argument, we can map every survival time $T^\prime_i$ following the distribution $S_0$ on a (random) survival time $T^{\prime\prime}_i\leq T^\prime_i$ such that $T^\prime\prime$ follows distribution $S_1$. For the tests with non-increasing scores, changing the realisation $\{T^\prime_i\}$ to $\{T^{\prime\prime}_i\}$ can only decrease the test statistic, leading to a larger p-value. This demonstrates strong alpha control.

For some other WLRT, however, the scores are not non-increasing. Take the example above where $w_j = 0$ for $t_j < 6$ and  $w_j = 1$ for $t_j \geq 6$. Here $c_j = 0$ for  $t_j < 6$ and $c_j$ is close to $1$ for the first $t_j$ greater than 6.

\subsection{Modestly weighted logrank tests (mWLRT)}

\begin{figure}[htbp]
\centering
\includegraphics[width=\linewidth]{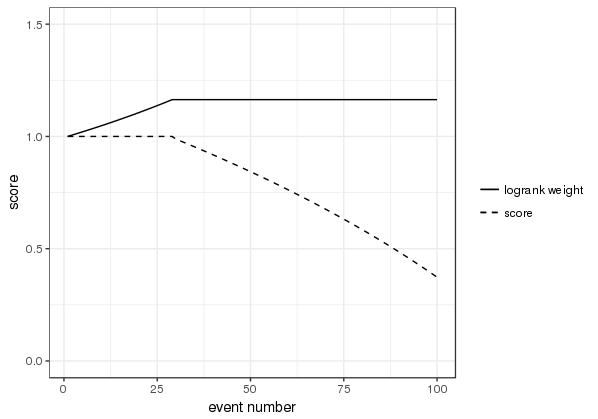}
\caption{Score-test and logrank-test representations of a modestly-weighted logrank test (mWLRT). Based on a 2-arm trial with $n =100$ per arm, where $100$ events are observed prior to any censored observations, with $30$ events observed prior to $t^*$, and with $100$ observations censored.}
\label{constant_scores}
\end{figure}

The heuristic argument above suggests a new class of WLRT that controls the type I error rate under $H_0^{\text{strong}}$. The idea is to start with the score statistic formulation of the WLRT and fix $c_j = 1$ for $t_j < t^*$. We can use (\ref{c_to_w}) to derive $w_1,\ldots,w_{j^*}$ and $C_1,\ldots,C_{j^*}$, where $t_{j*}:= \min t_j : t_j < t^*$. Thereafter, however, we fix $w_j =  w_{j^*}$ for $j = j^*+1,\ldots,k$ and use (\ref{w_to_c}) to derive $c_{j^* + 1},\ldots,c_{k}$ and $C_{j^* + 1},\ldots,C_{k}$. As an example, consider a 2-arm trial with $n =100$ per arm, where $100$ events are observed prior to any censored observations, with the remaining $100$ patients censored. Further suppose that $30$ out of the $100$ events are observed prior to  $t^{*}$. The resulting scores are shown in Figure \ref{constant_scores}. We call this class of tests "modestly weighted logrank tests (mWLRT)" because the range of logrank weights is far smaller than in previous proposals.

The larger $t^*$, the lower the relative weight given to early events, which is a good thing if we expect a delayed effect. On the other hand, if $t^*$ is too large then later events will get very high $w_j$, which may be inefficient. The choice of $t^*$ involves a trade-off of these two things. We can use simulations to guide us.

\section{Simulation study}

\subsection{Scenarios}

We consider a two-arm RCT with 100 patients per arm, recruited at a uniform rate over 12 months. On the control arm, survival times are exponentially distributed with median 15 months. Data cut-off is at a calendar time of 36 months after the start of the study, at which point any patients still alive are censored. The following four scenarios differ only according to the survival distribution on the experimental arm.
\begin{enumerate}[I.]
\item Weak null hypothesis: survival on the experimental arm is exponentially distributed with median 15 months, i.e., with rate $\log(2) / 15 = 0.046$ \\
\item Strong null hypothesis: survival on the experimental arm follows a 2-piece exponential distribution. For the first 6 months the rate is $\log(2) / 9 = 0.077$; thereafter the rate is $0.04$.\\
\item New treatment efficacious (PH): survival on the experimental arm is exponentially distributed with median 24 months, i.e., with rate $\log(2) / 24 = 0.029$\\
\item New treatment efficacious (NPH): survival on the experimental arm follows a 2-piece exponential distribution. For the first 6 months the rate is $\log(2) / 15 = 0.046$; thereafter the rate is $\log(2) / 30 = 0.023$.
\end{enumerate}

\begin{figure}[htbp]
\centering
\includegraphics[width=\linewidth]{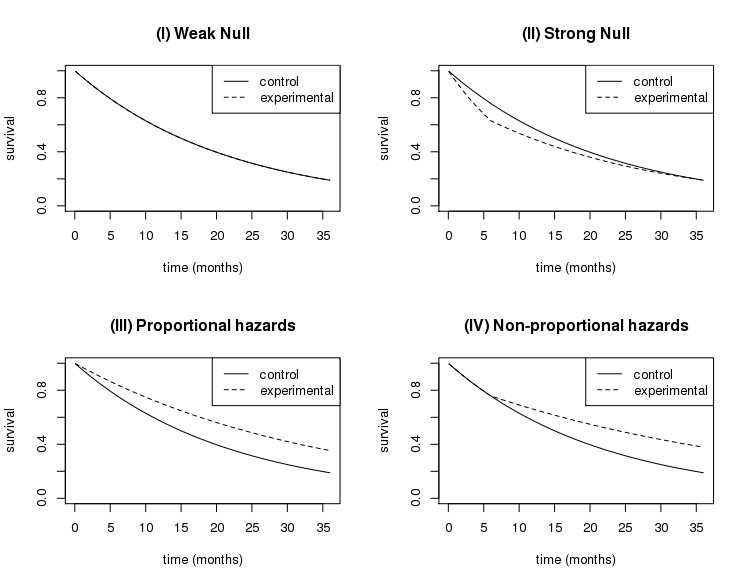}
\caption{Simulation study scenarios.}
\label{scenarios}
\end{figure}

The survival curves corresponding to these scenarios are shown in Figure \ref{scenarios}.

\subsection{Methods}

We compare three methods:

\begin{enumerate}
 \item $\text{WLRT}(t^*)$, where $t^* \in \left\lbrace 0,6 \right\rbrace$.
 \item $\text{mWLRT}(t^*)$, where $t^* \in \left[3,30\right]$.
 \item $\text{Landmark}(t^*)$, where $t^* \in \left[15,30\right]$.
\end{enumerate}



Here, $\text{WLRT}(t^*)$, refers to a weighted logrank test where $w_j=0$ if $t_j < t^*$ and $w_j = 1$ otherwise. So that $\text{WLRT}(0)$ is the standard logrank test.

A landmark analysis at time $t$ uses 
\begin{equation}
\frac{\hat{S}_1(t) - \hat{S}_0(t)}{\sqrt{\text{var}\left\lbrace\hat{S}_1(t)\right\rbrace+\text{var}\left\lbrace\hat{S}_0(t)\right\rbrace}} \nonumber
\end{equation}
as the standardized test statistic, based on Kaplan-Meier estimation.

\subsection{Results}

\begin{figure}[htbp]
\centering
\includegraphics[width=\linewidth]{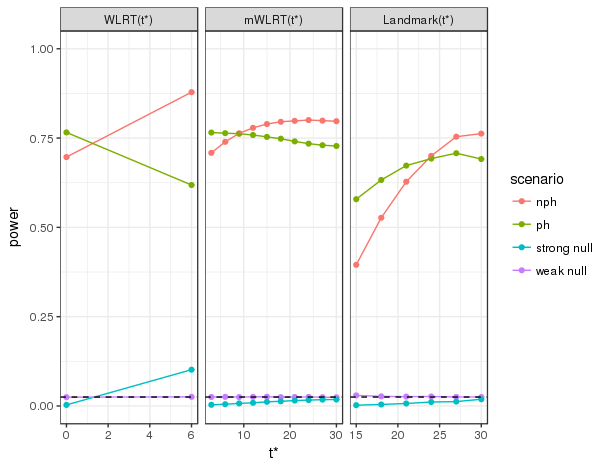}
\caption{Power and type 1 error of weighted logrank and landmark analyses under the four scenarios from Figure \ref{scenarios}.}
\label{power}
\end{figure}

Simulation results, based on 10,000 replications, are presented in Figure \ref{power}.

Under the weak null hypothesis $S_1(t) = S_0(t)$ all methods control the one-sided type I error rate at the nominal 0.025 level.

Under scenario II, where the survival curves are contained in $H_0^{\text{strong}}$, $\text{WLRT}(6)$ has $\text{pr}(U/\sqrt{V} > \Phi^{-1}(1 - 0.025))$ substantially greater than $0.025$. For all other methods, this probability is less than $0.025$.

Under scenario III, where the hazards are proportional, the standard logrank test has the highest power. However, the $\text{mWLRT}(t^*)$ are very close, suggesting that tests from this new class lose very little efficiency under proportional hazards. In contrast, $\text{WLRT}(6)$, which gives zero weight to early events, is substantially less powerful, as are the landmark analyses.

Under scenario IV it is $\text{WLRT}(6)$ that is optimal. However, the new tests do substantially better than the standard logrank test, and are competitive with the best landmark analysis.

We can also express these results in terms of relative efficiency. For example, compared to the standard logrank test, $\text{mWLRT}(18)$ has a relative efficiency of 

$$100 \times \left\lbrace \frac{\Phi^{-1}(0.975) + \Phi^{-1}(0.748)}{\Phi^{-1}(0.975) + \Phi^{-1}(0.766)} \right\rbrace = 96\%$$

under the proportional hazards scenario, and 

$$100 \times \left\lbrace \frac{\Phi^{-1}(0.975) + \Phi^{-1}(0.796)}{\Phi^{-1}(0.975) + \Phi^{-1}(0.697)} \right\rbrace = 127\%$$

under the nonproportional hazards scenario. In contrast, the best landmark analysis at $t^*=27$ has relative efficiency compared to the standard logrank test of $87\%$ under proportional hazards, and $114\%$ under nonproportional hazards.

\section{Discussion}

Evaluating a treatment is much more complicated than checking whether the p-value is less than $5\%$. In this section, we will discuss a number of topics related to the evaluation of drugs with potentially non-proportional hazards: 1) one-sided vs.\ two-sided testing; 2) Is rejecting the strong null hypothesis enough to show a real benefit?; 3) How can efficacy be estimated?; 4) Are there other promising tests?; 5) Further work.

In most standard situations, the relationship between one-sided and two-sided tests is trivial. Often, there is one key parameter of interest, e.g.\ cure rate, and if the null is rejected, it is obvious whether the conclusion should be that the new treatment has an efficacy advantage or disadvantage. This poses a risk that scientists do not recognise when one-sided testing is crucial. Although truly two-sided null hypotheses may be of interest in some other applications, they are virtually never of value for the primary analysis of confirmatory clinical trials. We have to show that the new drug has at least some efficacy advantage, not that it is different in an unspecified way. Our recommendation is therefore that regulatory guidance documents should consistently require one-sided hypothesis tests.

When time-to-event data are analysed, we are trying to compare two infinite-dimensional survival curves. Unless in a non-inferiority setting, we think that the minimal requirement for a new drug is that the strong null hypothesis of uniformly shorter (progression-free) survival can be rejected. Unfortunately, this is only concluding that survival is better at some time points; without further assumptions it is impossible to conclude that that all patients will benefit or that the survival rate will not be worse at some point in time. These issues are inherent in the experimental setting. 

We do not argue that the standard logrank test or Cox regression are invalid if hazards are non-proportional, but these tests have a clear benefit only if hazards are proportional or at least if that is a good approximation. As we have seen, the standard logrank test, which is often labelled "unweighted", does in fact weight observations differently and this particular weighting can lead to a very inefficient inference in the case of non-proportional hazards. Similarly, the estimated hazard ratio that comes with the standard methods only has a clear utility for (approximately) proportional hazards, and even then we still require supplementary measures. A hazard ratio of 0.5 has a different meaning clinically when median survival is 3 months compared to when it is 3 years. 

We argue that mWLRT have robust power when we expect a delayed separation of survival curves, as well as providing strong control of the type 1 error rate. For WLRT it is possible to construct an estimate of a weighted hazard ratio \citep{lin2017estimation}, but this will be equally, if not more, difficult to interpret as the standard hazard ratio estimate under non-proportional hazards. Therefore, additional measures will always be required. It is common to complement an analysis with estimated median survival. This is relatively easy to interpret, but ignores much of the information, and the full Kaplan-Meier curves will add information. The mean survival is clinically relevant and often a key parameter for reimbursement.

Tests based on restricted mean survival hold some promise, although censoring adds variability and the cutoff time for inclusion in the restricted mean is somewhat arbitrary. It may be useful to consider parametric approaches, tailored to likely survival distributions. We think that a sponsor should pre-specify the analysis and in that process try to predict trial outcome and uncertainties. With a fixed alternative hypothesis, it could in principle be possible to optimize the test across the different approaches that have been suggested. In practice, it is important also to consider the robustness of different inference options. Assurance, expected power over a distribution of scenarios, can be one useful operating characteristic.

\section*{Acknowledgements}

We thank Dr Jonathan Bartlett for helpful feedback on early versions of the manuscript.

\bibliographystyle{ama} 
\bibliography{nph} 

\end{document}